\documentclass{article}\usepackage{natbib,amssymb,psfig,emulateapj,epsfig}

\newcommand{\rem}[1]{ }
\newcommand{\beq}{\begin{equation}}
\newcommand{\eeq}{\end{equation}}
\newcommand{\bea}{\begin{eqnarray}}
\newcommand{\eea}{\end{eqnarray}}

\begin{document}

\title{Comment on ``Nonextensive theory of dark matter 
and gas density profiles''}

\author{Mikhail V. Medvedev\altaffilmark{1}}

\affil{Department of Physics and Astronomy, 
University of Kansas, KS 66045}
\altaffiltext{1}{Also: Institute for Nuclear Fusion, RRC ``Kurchatov
Institute'', Moscow 123182, Russia}

\begin{abstract}
Self-gravitating systems with nonlocal, long-range interactions
are described by nonextensive statistics.  Recently, Leubner
demonstrated that the nonextensivity parameter $\kappa$ should
be negative for self-gravitating, pressureless systems, such as
dark matter halos. The equation for the spherically symmetric 
nonextensive dark matter halos has also been derived. Here we demonstrate
that this equation is identical to the classical Lane-Emden equation
describing the structure of self-gravitating polytropic spheres. 
This establishes an intimate connection between self-gravitating 
polytropes and nonextensive thermostatistics. Moreover,
based on this fact and observational data, we put a stronger constraint on 
$\kappa$, namely $\kappa\la-3.4$. 
\end{abstract}

\keywords{cosmology: theory --- dark matter --- galaxies: halos
--- galaxies: structure}


\section*{}

The classical Boltzmann statistics describes systems with short-range
interactions and without long-term memory. In contrast, many astrophysical 
systems, e.g., self-gravitating objects,  
exhibit long-range interactions. Certain thermal properties, such as
(negative) heat capacity, can lead to long-term spatio-temporal correlations.
Therefore, the Boltzmann theory cannot, often, yield an adequate 
description of these systems. 

The generalization of the Boltzmann statistics proposed by \citet{T88} 
accounts for the systems in which the entropy is not an additive 
quantity. The generalized nonextensive entropy reads as \citep{L05}
\beq
S_\kappa=\kappa k_B\left(\sum p_i^{1-1/\kappa}-1\right),
\eeq
where $k_B$ is the Boltzmann constant, $p_i$ is the probability of 
the $i$-th microscopic state and $\kappa$ is the nonextensivity parameter,
also referred to as the entropic index. This parameter defines the 
``strength'' of long-range correlations in a system. This is a free parameter 
in the theory. For $\kappa\to\pm\infty$, one recovers the standard expression
for the entropy $S_\infty=-k_B\sum p_i\ln p_i$.

In a recent paper, \citet{L05} demonstrated that the nonextensive
statistics can be advantageous in describing cosmological self-gravitating
objects, e.g., dark matter halos of galaxies and clusters. 
One of the main results 
of that paper is Equation (6), which describes the spatial structure of
spherically symmetric, self-gravitating nonextensive particle systems. 
We note that, upon substitution $\psi=(\rho/\rho_0)^{1/n}$, this equation
can be cast into the form:
\beq
\frac{1}{r^2}\frac{d}{dr}\left(r^2\frac{d}{dr}\psi\right) 
=\frac{4\pi G\rho_0}{(3/2-n)\sigma^2}\,\psi^n,
\label{le}
\eeq
where $G$ is the gravitational constant, $\sigma$ and $\rho_0$ 
are the velocity dispersion and particle density normalization, and
$n=3/2-\kappa$. 

Equation (\ref{le}) is the classical Lane-Emden equation for $\kappa<0$, 
whereas it differs from the Lane-Emden equation by the sign of the 
right-hand-side for $\kappa>0$. The Lane-Emden equation describes
self-gravitating polytropic spheres. It has been studied in great details,
see, e.g., \citet{BT} for a general discussion and \citet{MR01} for recent 
advances. We emphasize that the established connection between the 
nonextensive statistics and self-gravitating polytropes is a very 
remarkable fact.

The entropic index $\kappa$ is related to the specific heat capacity 
of a system. Self-gravitating systems have negative heat capacity, 
hence the constraint \citep{L05}
\beq
\kappa<0,
\label{k<0}
\eeq
which corresponds to the index $n>3/2$. For such $\kappa$'s, Equation
(\ref{le}) has one (and only one) nontrivial exact 
solution\footnote{The other two integrable cases of the Lane-Emden equation,
$n=0$ and $n=1$, correspond to positive values of $\kappa$ and, 
hence are not of interest to us.}, namely for $n=5$ \citep{MR01}. Moreover,
for $n<5$ the polytropic spheres are finite in size. That is, the
particle density vanishes at a certain radius, $r_h$, --- the halo radius. 
Interestingly, for $n=5-\epsilon$ (with $\epsilon\ll1$), the ratio of the
halo radius to the core radius, $r_c$, can be calculated analytically 
\citep{MR01} to yield
\beq
\frac{r_h}{r_c}=\frac{32}{\pi\epsilon}.
\eeq 

The halo radius is not a well constrained observational parameter. 
We can put constraints on $r_h$ from the observed or derived halo 
parameters, such as the radius at which the X-ray brightness is 
detected at $>3\sigma$ level or the virial and/or tidal radii. 
A factor of few uncertainly introduced by this freedom does not 
induce much uncertainty in $n$, provided $r_h/r_c\gg1$.
Galactic halos have cores with $r_c$ of about one to few of kpc (with
a great diversity in the actual number). Observations cannot constrain
well the outer radii, which are likely greater than few tens of kpc.
This gives $r_h/r_c>10$, which yields constraints $n\ga4$ and, hence,
$\kappa\la-2.5$. Galaxy clusters provide a somewhat larger dynamical range
with $r_c$ being as small as few tens of kpc and the virial radii being
as large as few Mpc (see, e.g., \citealp{Vikhlinin+05}). The scales 
$r_h$ and $r_c$ are separated by 1.5 to two orders of magnitude, which yields
$n\ga4.9$. Hence, we put the observational constraint on the entropic index
\beq
\kappa\la-3.4\, .
\eeq
Further improvement in the $r_h/r_c$ ratio determination does not 
significantly affect this estimate, because $r_h/r_c\to\infty$ as 
$\kappa\to-3.5$. An interesting fact, not yet addressed by observations, 
is that for $n>5$, the dark matter profile has multiple cores \citep{MR01}. 
Provided that light traces mass, the cores may be observed
in the surface brightness profiles. The relations between
the core and break radii \citep{Vikhlinin+05} can set the 
value of $\kappa$ much more accurately.

\acknowledgements
This work is supported by
DoE grant DE-FG02-04ER54790, NASA grant NNG-04GM41G and the GRF fund.

 

\end{document}